\documentclass[a4paper,12pt]{article}
\usepackage{amsmath}
\usepackage{amssymb}

\DeclareMathOperator{\tr}{tr}
\def\lie{\pounds}
\def\cL{{\mathcal L}}
\def\cC{{\mathcal C}}

\begin{document}

\begin{flushright}
hep-th/0611141
\end{flushright}
\begin{center}
{\Large\bfseries 
Black Hole Entropy\\[1mm]
in the presence of Chern-Simons Terms}\\
\bigskip
 {\large Yuji Tachikawa}\\
\bigskip
\itshape
School of Natural Sciences, Institute for Advanced Study, \\
Princeton, New Jersey 08540, USA
\end{center}
\bigskip
\centerline{\large\textbf{abstract}}
\bigskip 

We derive a formula for the black hole entropy 
 in  theories with gravitational Chern-Simons terms,
by generalizing Wald's argument which uses the Noether charge.
It correctly reproduces the entropy 
of three-dimensional black holes in the presence of Chern-Simons term,
which was previously obtained via indirect methods.

\section{Introduction}
In the pursuit of the quantum theory of gravitation,
an important guidance comes from the laws of 
black hole thermodynamics, which can be established 
by semi-classical arguments.  Firstly,
manipulation of the classical equation of motion of general relativity yields
the first law of black hole dynamics \cite{BCH} which states \begin{equation}
\frac{\kappa}{2\pi} \frac{\delta A}{4 G_N}=\delta m -\Omega \delta J
\end{equation}where $A$ is the horizon area, $\kappa$ the surface gravity
at the horizon, $m$ the mass, $\Omega$ the angular velocity  and
$J$ the angular momentum of the black hole. $G_N$ is the Newton constant
entering the Lagrangian as \begin{equation}
\int d^Dx \sqrt{-g} \cL=\frac{1}{16\pi G_N}\int d^Dx \sqrt{-g}R\label{EH}
\end{equation} where $D$ is the number of
spacetime dimensions. The law shows a strong resemblance to the first
law of thermodynamics if we identify $\kappa$ with temperature
and $A$ with entropy. A key fact is that
semi-classical calculation \cite{HawkingRad}
tells us that in the black hole background
particles are produced in thermal ensemble at temperature \begin{equation}
T_H=\frac{\kappa}{2\pi}.
\end{equation} Thus it is very natural to set
the entropy of a black hole to be \begin{equation}
S=\frac{A}{4 G_N}\label{EHentropy}
\end{equation} and to search for its statistical explanation.

A remarkable success is that in superstring theory
a class of extremal charged black holes has a dual 
microscopic description using the D-branes
and that it reproduced the entropy \eqref{EHentropy}
in the large charge limit.
It also predicted subleading corrections which lead to a
deviation from the formula \eqref{EHentropy},
whose  existence
 was also expected from the macroscopic point of view because the low energy
limit of string theory includes higher derivative corrections to the 
Einstein-Hilbert Lagrangian \eqref{EH}.   Thankfully Wald and others
\cite{Wald,JKM,IyerWald} had already computed the general formula
of the black hole entropy of arbitrary general-covariant  Lagrangian
constructed from the metric, which is given by \begin{equation}
S=-2\pi\int_{\Sigma} d^{D-2}x \sqrt{-g}
\frac{\delta\cL}{\delta R_{\mu\nu\rho\sigma}}
\epsilon_{\mu\nu}\epsilon_{\rho\sigma}\label{Wald}.
\end{equation} Here $\Sigma$ is the horizon cross section and
$\epsilon_{\alpha\beta}$ is the binormal to the horizon\footnote{
It is defined as $\epsilon_{\alpha\beta}=n_\alpha \xi_\beta-\xi_\alpha n_\beta$
where $\xi_\alpha$ is the null Killing vector generating the horizon and
$n_\alpha$ is the 
ingoing null normal to the horizon cross section normalized to have 
$n_\alpha \xi^\alpha=-1$. }.
Another success of recent advances in quantum gravity
is that the deviation calculated from the application of Wald's formula \eqref{Wald}
to the superstring effective action agreed precisely with the
microscopic entropy calculated using the description as branes.
Interested readers can consult the review \cite{MoReview} for details.

A subtlety is that the compactification of superstring theory
down to odd dimensional spacetime often includes gravitational
Chern-Simons term, and in the case of three dimensions
it becomes the topologically massive gravity
first described by Deser, Jackiw and Templeton in \cite{DJT1,DJT2}.
It has solutions describing rotating black holes, first discovered by
Ba\~nados, Teitelboim and Zanelli  \cite{BTZ}, but the Wald's formula
\eqref{Wald} cannot be applied because
it is not manifestly invariant under diffeomorphism.
Thus various indirect methods were devised to treat such cases, e.g.
\cite{KrausLarsen,KrausLarsen2,SahooSen1,SahooSen2,JHC,MIP}.
The aim of this short note is to extend the argument in \cite{Wald}
to provide the entropy formula for black holes in the presence of
such Chern-Simons terms. We will see that the formula to be obtained
reproduces the correction to the entropy of three-dimensional black hole,
which was determined before.  Our formula is sufficiently general
so that  it can be applied for mixed gravitational Chern-Simons terms
and Green-Schwarz type couplings in any dimensions.

The rest of the paper is organized as follows. In the next section,
we will review the argument in \cite{Wald,JKM,IyerWald},
making the necessary changes along the way to accommodate
the gravitational Chern-Simons terms. At the end we will obtain the
generalized formula.  In section \ref{examples} we will
explicitly evaluate the formula for several examples,
and check that it agrees with the result in the literature.
We conclude with a short outlook in section~\ref{outlook}.

\section{General form of the entropy formula}\label{derivation}
We use a $D$-form $L(\phi)$ to represent the Lagrangian density,
where $\phi$ collectively denotes fundamental field variables.
A crucial property used in the derivation of the first law in \cite{Wald}
is that it satisfies the condition \begin{equation}
\delta_\xi L(\phi)=\lie_\xi L(\phi), \label{diffeo-inv}
\end{equation} 
where $\delta_\xi$ is the variation induced by the diffeomorphism by 
the vector field $\xi$,
while $\lie_\xi$ denotes the Lie derivative with respect to $\xi$.
The condition \eqref{diffeo-inv} is not satisfied for the gravitational
Chern-Simons term $L_{CS}$
 \begin{equation}
L_{CS}\sim \tr \Gamma \wedge R^{2m-1} +\cdots
\end{equation}  where $\Gamma$ is the affine connection,
$R$ the curvature two-form and $D=4m-1$. It satisfies, however, the 
 condition \begin{equation}
\delta_\xi L(\phi)=\lie_\xi L(\phi)+d\, \Xi_\xi.\label{mod-diffeo}
\end{equation} for a suitable $(D-1)$-form $\Xi_\xi$,
which is enough in obtaining generally covariant
equations of motion. Here and in the following we 
define the Lie derivative $\lie_\xi$ of various
non-tensor quantity e.g. $\Gamma$ to be the resulting expression 
if the quantity involved were tensorial with the same index structure. 
We will see below how the first law can be derived starting from the
generalized transformation law \eqref{mod-diffeo}, without assuming
the covariance in the strict sense, \eqref{diffeo-inv}.

First we express the first-order variation  of $L$ as \begin{equation}
\delta L=\sum_\phi E_\phi\delta\phi + d\Theta(\phi,\delta\phi),
\end{equation} 
 $E_\phi$ gives the equation of motion associated with $\phi$,
while $\Theta$ is called the symplectic potential.  
Let us set \begin{equation}
\Omega(\phi,\delta_1\phi,\delta_2\phi)=\delta_1\Theta(\phi,\delta_2\phi)
-\delta_2\Theta(\phi,\delta_1\phi).\label{symplecticform}
\end{equation} It is important that 
the integral of $\Omega$ over a Cauchy surface $\cC$, properly projected
down to the gauge-invariant phase space, defines
the symplectic form in the covariant phase space approach \cite{crn,LeeWald}.

Secondly, the condition \eqref{mod-diffeo} means that the diffeomorphism
generated by $\xi$ is a symmetry in the sense of \cite{LeeWald}, 
thus the corresponding current
$j_\xi$ given by
\begin{equation}
j_\xi=\Theta(\phi, \delta_\xi\phi)-\iota_\xi L-\Xi_\xi
\end{equation}  is conserved on-shell, i.e. $dj_\xi\simeq 0$.
Here $\simeq$ denotes the equality which holds only if the equation of motion
is satisfied, and $\iota_\xi$ is the interior  product.
Thus, from the analysis in \cite{WaldIdent}, $j_\xi$ is exact on-shell, i.e.
there is a $(D-2)$-form $Q_\xi$ constructed from the products of 
repeated derivatives of $\xi$ and $\phi$ such that
\begin{equation}
j_\xi\simeq dQ_\xi\label{Q}.
\end{equation}
As described in \cite{WaldIdent},
$Q_\xi$ can be constructed algorithmically given the form of $j_\xi$
independently of the detailed structure of the equations of motion.

Before going to the third step, let us define $\Pi_\xi$ by the equation\begin{equation}
\delta_\xi\Theta=\lie_\xi \Theta+\Pi_\xi.
\end{equation} Calculating $\delta\delta_\xi L$ in two ways, we obtain \begin{equation}
d\Pi_\xi \simeq \delta d\, \Xi_\xi.
\end{equation} Thus, using the theorem in \cite{WaldIdent} again, 
there exists a $(D-2)$-form $\Sigma_\xi$ such that \begin{equation}
\Pi_\xi-\delta\,\Xi_\xi\simeq d\Sigma_\xi.
\end{equation}

Then the third step is the following manipulation of the equation: \begin{align}
\delta j_\xi&= \delta \Theta(\phi,\delta_\xi \phi)-\iota_\xi \delta L 
-\delta\, \Xi_\xi \nonumber\\
&\simeq \delta \Theta(\phi,\delta_\xi\phi) -\lie_\xi \Theta(\phi,\delta\phi)
+d\iota_\xi \Theta -\delta\, \Xi_\xi \nonumber\\
&= \delta \Theta(\phi,\delta_\xi\phi) -\delta_\xi \Theta(\phi,\delta\phi)
+d\iota_\xi \Theta +\Pi_\xi-\delta\, \Xi_\xi \nonumber\\
&\simeq \Omega(\phi,\delta\phi, \delta_\xi\phi) 
+d(\iota_\xi \Theta +\Sigma_\xi).
\end{align} We need to find a quantity $C_\xi$ which satisfies \begin{equation}
\delta C_\xi =\iota_\xi \Theta+\Sigma_\xi,
\end{equation} and let us define \begin{equation}
Q'_\xi = Q_\xi-C_\xi.\label{Qprime}
\end{equation} Then we finally
 arrive at the following relation \begin{equation}
\delta dQ'_\xi\simeq \Omega(\phi,\delta\phi,\delta_\xi \phi),\label{ham}
\end{equation} which means that $dQ'_\xi$ is the Hamiltonian
generating the diffeomorphism $\xi$
in the covariant phase space approach.

Suppose we have a stationary black hole spacetime
with a bifurcate Killing horizon\footnote{A bifurcate horizon
is a pair of Killing horizons generated by the same Killing vector $\xi$
intersecting on a spacelike $(D-2)$-dimensional surface $\Sigma$,
on which $\xi$ vanishes. $\Sigma$ is called the bifurcation surface.
A most simple example is that of the Schwarzschild black hole in the Kruskal
coordinate. The bifurcation surface is at its origin, and the Killing vector
$\xi$ acts as the Lorentz boost with strength $\kappa$ near the origin.

Note that in the 
static coordinate system, any spacelike surface at constant $t$ passes $\Sigma$.
Thus the horizon cross section at any finite $t$
\textit{is} the bifurcation surface.
Physically sensible spacetime with stationary  black holes with nonzero 
constant $\kappa$ is expected
to have an  extension in which the horizon becomes bifurcate.
For more discussion, we refer the reader to  \cite{bifurcate}
and the references therein. 
} generated by $\xi$ such that $\delta_\xi \phi=0$.
The right hand side of \eqref{ham} then vanishes.
Suppose $\xi=t + \Omega \phi $ where $t$ is the generator of the
global time translation, $\Omega$ the angular velocity of the
horizon and $\phi$ the angular rotation. 
Integrating \eqref{ham} on a Cauchy surface,
one obtains
\begin{equation}
\delta\int_\Sigma Q'_\xi  \simeq
 \delta\int_\infty Q'_t +\Omega \, \delta \int_\infty Q'_\phi.
\label{pre-1st-law}
\end{equation}  where $\Sigma$ is the horizon cross section
and $\infty $ the asymptotic infinity of the Cauchy surface. 
 Since $dQ'_\eta$  is the Hamiltonian for the diffeomorphism $\eta$,
$E=\int_\infty Q'_t$  is the energy
which includes the ADM mass and contribution from
long range gauge forces,
and $J=-\int_\infty Q'_\phi$ is the angular momentum.

In evaluating the left hand side of \eqref{pre-1st-law},
 repeated derivative of $\xi$ can be traded
with Riemann curvature times lower derivative of $\xi$ using the Killing identity,
thus $dQ'$ can be made to be  a linear combination of $\xi$ and $\nabla_a \xi_b$.
It is known \cite{JKM} that we can assume that $\Sigma$ is the bifurcation surface
without changing the integral, and thus  $\xi=0$ and
$\nabla_a\xi_b=\kappa \epsilon_{ab}$  on $\Sigma$.
Let us define, then,
\begin{equation}
S=2\pi \int_\Sigma Q'_\xi \big|_{\xi\to 0, \ \nabla_a\xi_b\to \epsilon_{ab}}.
\end{equation}  Now the equation \eqref{pre-1st-law} becomes
the first law which states: \begin{equation}
\kappa \delta S\simeq \delta E-\Omega \delta J.
\end{equation}

The algorithm which determines $Q'$ from $L$ is linear.
Thus in finding the contribution to the entropy from
the Chern-Simons term, we can follow the procedure described above 
for $L$ being solely given by the Chern-Simons term, and add the result
to the contribution from covariant terms. 
Unfortunately we have not been able to obtain 
a general formula similar to the celebrated Wald's formula \eqref{Wald}. 
We will analyze them on a case-by-case basis
in the next section. 

As we saw, our algorithm depends on finding $\Sigma_\xi$ etc., which has
an arbitrariness of the form $\Sigma_\xi\to \Sigma_\xi +dX$. 
Before continuing,  let us check that these ambiguities do not change
the entropy. Firstly the shift just mentioned changes $Q'_\xi$ by an
exact form, which does not modify its integral over the horizon cross
section.  Secondly, the change of $\Xi_\xi$  to $\Xi_\xi + d Y$ induces
$j\to j-\delta dY$,
while it changes $\Sigma_\xi $ to $\Sigma -\delta Y$. Thus $Q'_\xi$ is left
intact.  Thirdly, the change in $Q'_\xi$ caused by  $\Theta \to \Theta + dZ $ 
can be seen to be proportional to $\delta_\xi \phi$ or $\xi$,
which vanishes on $\Sigma$.  Finally, the addition to the Lagrangian
of a total derivative term $L\to L+dW$ results in the shift
$\Theta\to \Theta+\delta W$ and $\Xi_\xi \to \Xi_\xi +\delta_\xi W$.
Thus it does not change $j_\xi$ or  $\Sigma_\xi$, which means
$Q'_\xi$ is unchanged.

\section{Examples}\label{examples}
\subsection{Three-dimensional gravitational Chern-Simons}

Firstly let us consider three-dimensional gravity with Chern-Simons
term \begin{equation}
L_{CS}=\beta\tr (\Gamma R-\frac13\Gamma^3),
\end{equation} where the wedge product should be understood.
Since \begin{equation}
(\delta_\xi-\lie_\xi) L_{CS}=\beta \tr d U_\xi\cdot d \Gamma
\end{equation} where $U_\xi$ is the matrix with component $(U_\xi)^\mu_\nu=\xi^\mu_{,\nu}$,
we can choose $\Xi_\xi=-\beta\tr dU \Gamma$.
Non-covariant contribution to the symplectic potential 
$\Theta$ is given by  $-\beta\tr \Gamma \delta \Gamma$, hence
$\Pi_\xi=-\beta\tr dU_\xi \delta \Gamma$. Thus for this choice $\Sigma_\xi=0$ 
and \begin{equation}
j_\xi= 2 \beta\tr dU_\xi \Gamma + 
\text{ (terms linear in $\nabla \xi $ and $\xi$)}.
\end{equation} Thus \begin{equation}
Q'_\xi=  2\beta \tr U_\xi \Gamma +\text{ (terms linear in $\xi$)}.
\end{equation}
Evaluating $2\pi\int Q'|_{\xi\to 0,\nabla_a \xi_b\to \epsilon_{ab}}$ 
on the bifurcation surface, we arrive at the following formula
\begin{equation}
\Delta S_{CS}= 8\pi\beta \int_\Sigma \Gamma_N\label{1d-cs}
\end{equation} where\footnote{In our convention,
the signature of the metric is mostly plus, and $\epsilon_{01\cdots}$ is positive.
The binormal is defined so that $\epsilon_{01}$ is positive. Then $\epsilon^1{}_0$ and $\epsilon^0{}_1$ are negative.} \begin{equation}
\Gamma_N=-\epsilon^\nu{}_\mu \Gamma^\mu{}_{\nu\rho} dx^\rho/2\label{21}
\end{equation}is the projection of the affine connection
to the normal bundle of $\Sigma$.

An important fact here is  that the binormal $\epsilon_{ab}$ is covariantly constant
on the bifurcation surface, because \begin{equation}
\nabla_c \epsilon_{ab}=\kappa^{-1} \nabla_c\nabla_a \xi_b = \kappa^{-1}
R_{abcd}\xi^d=0.
\end{equation}  It means that the holonomy of the metric on the bifurcation surface 
is reduced to lie in $SO(1,1)_N\times SO(D-2)_T$ where
$N$, $T$ stands for normal and tangential, respectively.
The factor $\epsilon^\mu{}_\nu$ in \eqref{21} projects the connection
to the normal component. Then  $\int_\Sigma \Gamma_N$ is the 
one-dimensional Chern-Simons term for the $SO(1,1)$
 connection of the normal bundle, and thus is gauge-invariant.

The formula \eqref{1d-cs} agrees with the formula obtained in \cite{Solo} using
the conical singularity method.
For the rotating Ba\~nados-Teitelboim-Zanelli black hole \cite{BTZ} with the metric 
\begin{multline}
-\frac{(\rho^2-\rho_+^2)(\rho^2-\rho_-^2)}{l^2\rho^2}d\tau^2\\
+\frac{l^2\rho^2}{(\rho^2-\rho_+^2)(\rho^2-\rho_-^2)}d\rho^2
+\rho^2(dy-\frac{\rho_+\rho_-}{l\rho^2}d\tau)^2
\end{multline}which has the outer horizon at $\rho=\rho_+$,
the contribution is \begin{equation}
\Delta S=8\pi\beta \int_\Sigma \frac{-\rho_+\rho_-}{l r}=-\frac{16\pi^2\beta\rho_-}l.
\end{equation} 
It agrees with the result given in the literature. The methods employed
were diverse:  it was done
in \cite{KrausLarsen,KrausLarsen2}
via the consideration of the boundary Virasoro algebra, 
in \cite{SahooSen1} by dimensional reduction to two dimensions 
and application of the Wald's formula, 
in \cite{JHC} using the description of
three-dimensional gravity as the $SO(2,2)$ Chern-Simons theory,
and in \cite{MIP} by the direct  integration of  the first law.

\subsection{Gravitational Chern-Simons in other dimensions}

We can generalize the formula in the previous subsection
to the following mixed
 gravitational Chern-Simons term \begin{equation}
L_{CS}=(\tr \Gamma R^{2m-1} + \cdots )  P(F) \label{genCS}
\end{equation} where $P(F)$ is a closed form constructed out of
fields other than the metric and $\cdots$ is determined from the descent 
relation $dL_{CS}=(\tr R^{2m})P(F)$. 
The calculation which lead to \eqref{1d-cs}
can be repeated almost verbatim  to yield \begin{equation}
\Delta S= 8\pi m\int_\Sigma \Gamma_N  R_N^{2m-2} P(F),\label{ggg}
\end{equation}  where $R_N$ is the curvature two-form for the normal bundle.
 All of the would-be contribution from
the ellipsis $\cdots$ in \eqref{genCS} vanishes since $SO(1,1)_N$ is Abelian.

If there is a form $\omega$ such that $P(F)=d\omega $, the result \eqref{ggg}
can be inferred from the original Wald's formula \eqref{Wald}. Indeed,
The term \eqref{genCS} then  has the same effect with the term \begin{equation}
L_{CS}'= \tr  R^{2m} \wedge \omega,\label{genCSp}
\end{equation} where  the affine connection does not appear explicitly.
Thus we can apply the formula \eqref{Wald} to obtain the contribution
to the entropy \begin{equation}
\Delta S=8\pi m \int_\Sigma  R_N^{2m-1} \wedge \omega,
\end{equation} which is equivalent to the equation \eqref{ggg}.
Strictly speaking, $\omega$ needs to have a non-trivial gauge transformation
between the coordinate patches on $\Sigma$ if $P(F)$ has non-trivial
flux through $\Sigma$. Thus one cannot naively partially integrate
from \eqref{genCS} to \eqref{genCSp} and one needs to be more careful. 
It would be interesting to rederive our formula using an auxiliary spacetime
which has one extra dimension as is usually done for the correct
definition of Chern-Simons terms in the presence of non-trivial fluxes.

\subsection{Lagrangian of Green-Schwarz type}

As another application of our analysis, let us determine
the contribution to the entropy from the term in the
six-dimensional Lagrangian 
of the form \begin{equation}
L= \frac 1{2g^2} H\wedge *H \label{GS}
\end{equation} where $H=dB+\lambda \tr(\Gamma R-\Gamma^3/3)$.
In order to maintain the general covariance,
the $B$ field needs to have an extra term in the transformation
under diffeomorphism given so that   \begin{equation}
\delta_\xi B= \lie_\xi B - \lambda \tr  dU_\xi \Gamma, 
\end{equation}where $U_\xi$ is 
the matrix $(U_\xi)^\mu_{\nu}=\xi^\mu_{,\nu}$ as before.
Modified transformation law of the form-fields such as this
often occurs in string theory in order to cancel the anomaly \`a la
Green-Schwarz mechanism.

Now $\Xi_\xi=0$ since $L$ itself is covariant. The non-covariant portion
of  $\Theta$ is \begin{equation}
\frac1{g^2} (\delta B-\lambda \tr \Gamma \delta \Gamma )\wedge *H
\end{equation}  and thus  $\Pi_\xi=-\lambda \tr dU \delta \Gamma/g^2$.
Plugging into the formula for $Q'$ \eqref{Qprime}, we obtain \begin{equation}
Q'_\xi=\frac{2}{g^2}\lambda \tr U_\xi \Gamma + (\text{terms linear in $\xi$}).
\end{equation}
Hence the contribution
 to the entropy is given by \begin{equation}
\Delta S=8\pi \frac{ \lambda}{g^2} \int_\Sigma \Gamma_N \wedge *H.\label{boo}
\end{equation}  

One can also determine the same contribution, of course, by the use 
of the dualization and the formula \eqref{ggg}
derived in the previous subsection.
Indeed, we can dualize the term \eqref{GS} to the form\begin{equation}
L=\frac{g^2}2 H_D \wedge *H_D -
H_D\wedge \lambda \tr( \Gamma R - \frac13\Gamma^3),
\end{equation}where $H_D=dB_D$ and $H_D=*H/g^2$ on-shell. Applying
the formula \eqref{ggg} we obtain \begin{equation}
\Delta S=8\pi \lambda \int_\Sigma \Gamma_N \wedge H_D,\label{bar}
\end{equation} which agrees with \eqref{boo} as it should be.

The formula obtained above can be checked against
the entropy  of the dyonic black hole
treated in \cite{SahooSen2}, where the entropy
was determined by means of dimensional  reduction.
There, the Lagrangian of the theory contains
a two-form with the kinetic term given by
\eqref{GS} and the theory is compactified
on $T^2$. The black hole has a two-dimensional horizon cross section
from the four-dimensional point of view, and one of the direction
of $T^2$ is fibered non-trivially on it. Thus, the black hole
has
a  four-dimensional horizon cross 
section of the topology $S^3\times S^1$ if viewed as a six-dimensional
spacetime.
One can then evaluate the correction to the entropy
using the expression \eqref{boo} or \eqref{bar}
on the background, which was given in (3.4) of \cite{SahooSen2}.
$H_D$ has an integral flux through the $S^3$ part of the horizon,
and then the correction reduces to the calculation of the one-dimensional
Chern-Simons term on $S^1$.  
It reproduces the first term in (3.34) of \cite{SahooSen2}. The second term
in (3.34) arose in \cite{SahooSen2}
from the modification of the background geometry by the Chern-Simons interaction.
In order to reproduce it from our formula \eqref{boo} we will first
need to determine the backreaction of the Chern-Simons term, which is left
to a future work.

\section{Summary and Outlook}\label{outlook}

In this short note, we studied  how one can generalize the argument in 
\cite{Wald,JKM,IyerWald} to obtain the entropy formula for the black holes 
in the presence of Chern-Simons terms. We gave an explicit formula
for several kinds of Chern-Simons terms.
It correctly reproduced  the contribution of
the three-dimensional gravitational Chern-Simons  term to the entropy,
which had already been determined via other methods.

Concrete examples we saw above were both related
to the BTZ black hole background, because the dyonic black hole in 
six dimensions reduces to the BTZ black hole if one treats $S^3$ part of the
horizon as the internal space.
Thus, it would be interesting to study the effect of Chern-Simons
terms in other dimensionality, say, $L_{CS}=\tr \Gamma R^3+\cdots$
for black holes in seven dimensions. 
Some string/M-theory compactification is known to have such interaction
in seven dimensions, thus it would be interesting to consider 
 brane construction for such black holes and check if the microscopic
entropy matches with the macroscopic prediction here.

Another point is that
Wald's formula and our generalized ones
allow us to calculate the correction to the entropy from
higher derivative terms
only after we know the correction to the background from the
same higher derivative terms.
The entropy function formalism \cite{Sen} 
devised by Sen is more convenient in that respect
because it automatically
incorporates both sources of corrections via the extremalization
of the entropy function.  Thus 
an interesting direction of research will be  to generalize the said
formalism to Lagrangians containing
Chern-Simons terms. 

\bigskip
\section*{Acknowledgements}
The work emerged from the conversation with Kentaro Hanaki.
The author greatly  benefited also from the discussion with
Masaki Shigemori. 
He would like to thank for the hospitality
of the people at the Aspen Center for Physics, where this work was initiated.
He was supported by the JSPS Research Fellowships
for Young Scientists at the initial stage of the work.
He is now supported by DOE grant DE-FG02-90ER40542.

\end{document}